\documentclass[twocolumn,superscriptaddress,aps,showpacs,amsmath,footinbib,bibnotes]{revtex4-1}
\pdfoutput=1
\usepackage{graphicx}
\usepackage{amssymb,amsmath}
\usepackage{textcomp} 
\usepackage[bold]{hhtensor}
\usepackage{natbib}
\usepackage{lipsum}
\usepackage{bm}
\usepackage{hyperref}
\usepackage{multirow}
\usepackage{float}
\usepackage{amsmath}
\usepackage{amssymb}
\usepackage{stmaryrd}
\usepackage{mathrsfs}
\usepackage[dvipsnames]{xcolor}
\hypersetup{
	colorlinks,
	linkcolor={red!80!black},
	citecolor={Blue},
	urlcolor={MidnightBlue}
}

\begin{document}

\author{Mian Zhang}
\thanks{These authors contributed equally to this work}
\affiliation{John A. Paulson School of Engineering and Applied Sciences, Harvard University, Cambridge, Massachusetts 02138, USA}
\author{Brandon Buscaino}
\thanks{These authors contributed equally to this work}
\affiliation{Edward L. Ginzton Laboratory, Department of Electrical Engineering, Stanford University, Stanford, California 94305, USA}
\author{Cheng Wang}
\thanks{These authors contributed equally to this work}
\affiliation{John A. Paulson School of Engineering and Applied Sciences, Harvard University, Cambridge, Massachusetts 02138, USA}
\affiliation{Department of Electronic Engineering, City University of Hong Kong, Kowloon, Hong Kong, China}
\author{Amirhassan Shams-Ansari}
\affiliation{John A. Paulson School of Engineering and Applied Sciences, Harvard University, Cambridge, Massachusetts 02138, USA}
\affiliation{Department of Electrical Engineering and Computer Science, Howard University, Washington DC 20059, USA}
\author{Christian Reimer}
\affiliation{John A. Paulson School of Engineering and Applied Sciences, Harvard University, Cambridge, Massachusetts 02138, USA}
\author{Rongrong Zhu}
\affiliation{John A. Paulson School of Engineering and Applied Sciences, Harvard University, Cambridge, Massachusetts 02138, USA}
\affiliation{The Electromagnetics Academy at Zhejiang University, College of Information Science and Electronic Engineering, Zhejiang University, Hangzhou 310027, China}
\author{Joseph M. Kahn}\email{jmk@ee.stanford.edu}
\affiliation{Edward L. Ginzton Laboratory, Department of Electrical Engineering, Stanford University, Stanford, California 94305, USA}
\author{Marko Loncar}\email{loncar@seas.harvard.edu}
\affiliation{John A. Paulson School of Engineering and Applied Sciences, Harvard University, Cambridge, Massachusetts 02138, USA}

\date{\today}
\title{Broadband electro-optic frequency comb generation in an integrated microring resonator}

\begin{abstract}
Optical frequency combs consist of equally spaced discrete optical frequency components and are essential tools for optical communications, precision metrology, timing and spectroscopy. To date, wide-spanning combs are most often generated by mode-locked lasers or dispersion-engineered resonators with third-order Kerr nonlinearity. An alternative comb generation method uses electro-optic (EO) phase modulation in a resonator with strong second-order nonlinearity, resulting in combs with excellent stability and controllability. Previous EO combs, however, have been limited to narrow widths by a weak EO interaction strength and a lack of dispersion engineering in free-space systems. In this work, we overcome these limitations by realizing an integrated EO comb generator in a thin-film lithium niobate photonic platform that features a large electro-optic response, ultra-low optical loss and highly co-localized microwave and optical fields, while enabling dispersion engineering. Our measured EO frequency comb spans more than the entire telecommunications L-band (over 900 comb lines spaced at $\sim$ 10 GHz), and we show that future dispersion engineering can enable octave-spanning combs. Furthermore, we demonstrate the high tolerance of our comb generator to modulation frequency detuning, with frequency spacing finely controllable over seven orders of magnitude (10 Hz to 100 MHz), and utilize this feature to generate dual frequency combs in a single resonator. Our results show that integrated EO comb generators, capable of generating wide and stable comb spectra, are a powerful complement to integrated Kerr combs, enabling applications ranging from spectroscopy to optical communications. 
\end{abstract}
\maketitle

The migration of optical frequency comb generators to integrated devices is motivated by a desire for efficient, compact, robust, and high repetition-rate combs \cite{kippenberg_microresonator-based_2011,ye_optical_2003}. At present, almost all on-chip frequency comb generators rely on the Kerr (third-order, $\chi^{(3)}$) nonlinear optical process, where a continuous wave (CW) laser source excites a low-loss optical microresonator having a large Kerr nonlinear coefficient. This approach has enabled demonstration of wide-spanning Kerr frequency combs from the near- to mid-infrared in many material platforms \cite{delhaye_octave_2011,okawachi_octave-spanning_2011,griffith_silicon-chip_2015,savchenkov_tunable_2008,liang_generation_2011}. Owing to the complex nature of the parametric oscillation process, however, the formation dynamics and noise properties of the Kerr combs are not yet fully understood and are still under active investigation \cite{herr_universal_2012,yi_single-mode_2017}. Sophisticated control protocols are typically required to keep Kerr combs stabilized.

\begin{figure*}
	\centering
	\includegraphics[angle=0,width=0.7\textwidth]{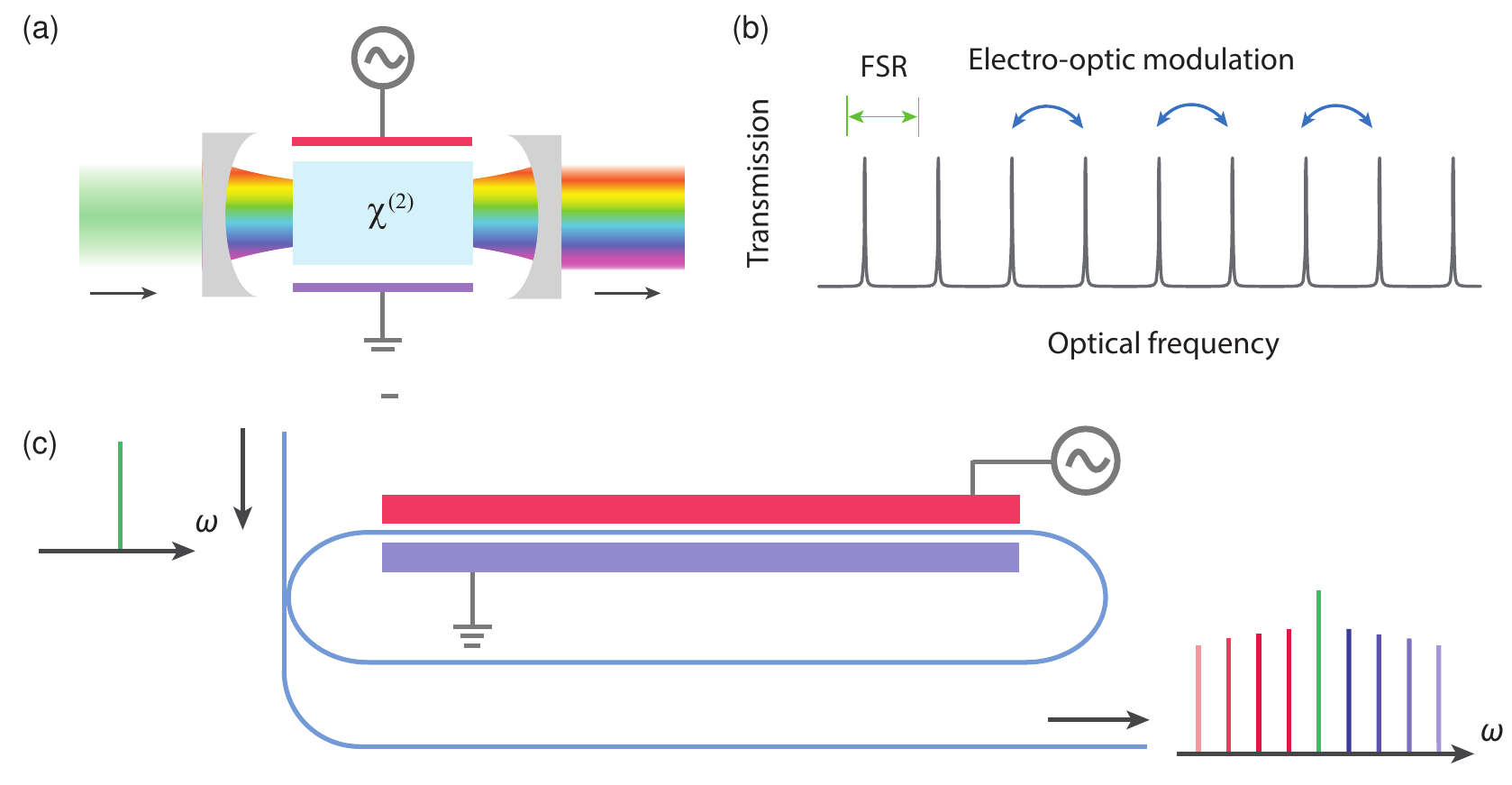}
	
	\caption{\label{fig1}\textbf{Resonator-enhanced electro-optic comb generator.}\textbf{(a)} Schematic of a canonical electro-optic (EO) comb generator comprising an EO ($\chi^{(2)}$) phase modulator inside a free-space Fabry-Perot (FP) resonator. A continuous-wave (CW) laser is coupled into the resonator and an optical frequency comb is generated at the output. \textbf{(b)} EO comb generation principle. A microwave signal, with modulation frequency equal to the free spectral range (FSR) of the optical resonator, couples light between different resonator modes. As a result, the input-coupled CW light is modulated, giving rise to sidebands at the modulation frequency, which are then recirculated to be modulated again. The modulation index determines the strength of coupling between nearby frequency components after passing through the modulator. \textbf{(c)} Integrated microring EO comb generator. The FP resonator can be replaced by a microring resonator that is EO modulated at a frequency matching the FSR of the ring. Similar to the FP resonator, a CW laser coupled into the ring resonator will be converted to a frequency comb in the output optical waveguide. }
\end{figure*}

An alternative frequency comb-generation method uses the electro-optic (EO) effect in materials with second-order ($\chi^{(2)}$) nonlinearity. Conventionally, EO frequency comb generators pass a CW laser through a sequence of discrete phase and amplitude modulators \cite{metcalf_high-power_2013,xiao_toward_2008,beha_electronic_2017}. Such EO comb generators can feature remarkable comb power and flat spectra, and can support flexible frequency spacing. They usually have narrow bandwidth, however, comprising only tens of lines and spanning only a few nanometers \cite{metcalf_high-power_2013,sakamoto_asymptotic_2007,ozharar_ultraflat_2008}. Therefore, highly nonlinear fiber is typically required to further broaden the comb spectrum, increasing the system complexity and size \cite{beha_electronic_2017}. 

Broader EO combs can be generated using an optical resonator to increase the nonlinear interaction strength \cite{kourogi_wide-span_1993,ho_optical_1993}. In a canonical resonator-based EO comb generator, a CW laser is coupled to a bulk nonlinear crystal resonator containing an EO phase modulator (Fig. \ref{fig1}a), and comb lines are generated solely through the $\chi^{(2)}$ process. When the modulation frequency matches a harmonic of the resonator free spectral range (FSR), the optical sidebands generated by the phase modulator are resonant. In a low-loss resonator, the light passes through the modulator many times before being dissipated or coupled out, efficiently generating many comb lines spaced at the modulation frequency (Fig. \ref{fig1}b). The output frequency comb can be predicted accurately by closed-form solutions \cite{ho_optical_1993} with spacings equal to the modulation frequency. The overall flatness of the comb strongly depends on the round-trip modulation strength and the optical resonator loss. In particular, at frequencies away from the pump frequency, the comb line power decreases exponentially: the optical power in the $q$th comb line is $P_q \propto e^{-|q|l/\beta}$, where $\beta=V_p/V_\pi$ is the phase modulation index, $V_p$ is the microwave drive peak amplitude, $V_\pi$ is the half-wave voltage of the phase modulator, $l=\frac{\pi\kappa}{\textrm{FSR}}$ is the round-trip electric-field loss coefficient of a resonator with damping rate $\kappa=\omega_0/Q$, $Q$ is the resonator quality factor, and $\omega_0$ is the optical frequency. It is therefore clear that strong phase modulation (large  $\beta$) and a high-$Q$ optical resonator (small $l$) are crucial for generating flat and broad EO combs. Furthermore, dispersion sets a fundamental limit on the total comb bandwidth by introducing frequency-dependent phase shifts that cause comb lines far from the pump frequency to fall out of resonance (see Supplementary Information). Although EO frequency combs generated by free-space or fiber-based optical cavities have been designed and extensively studied for over 25 years \cite{kourogi_wide-span_1993,ho_optical_1993,kourogi_limit_1995}, practical comb widths are still limited to a few tens of nanometers by a combination of weak modulation and limited dispersion engineering \cite{kourogi_limit_1995}. 

\begin{figure*}
	\centering
	\includegraphics[angle=0,width=\textwidth]{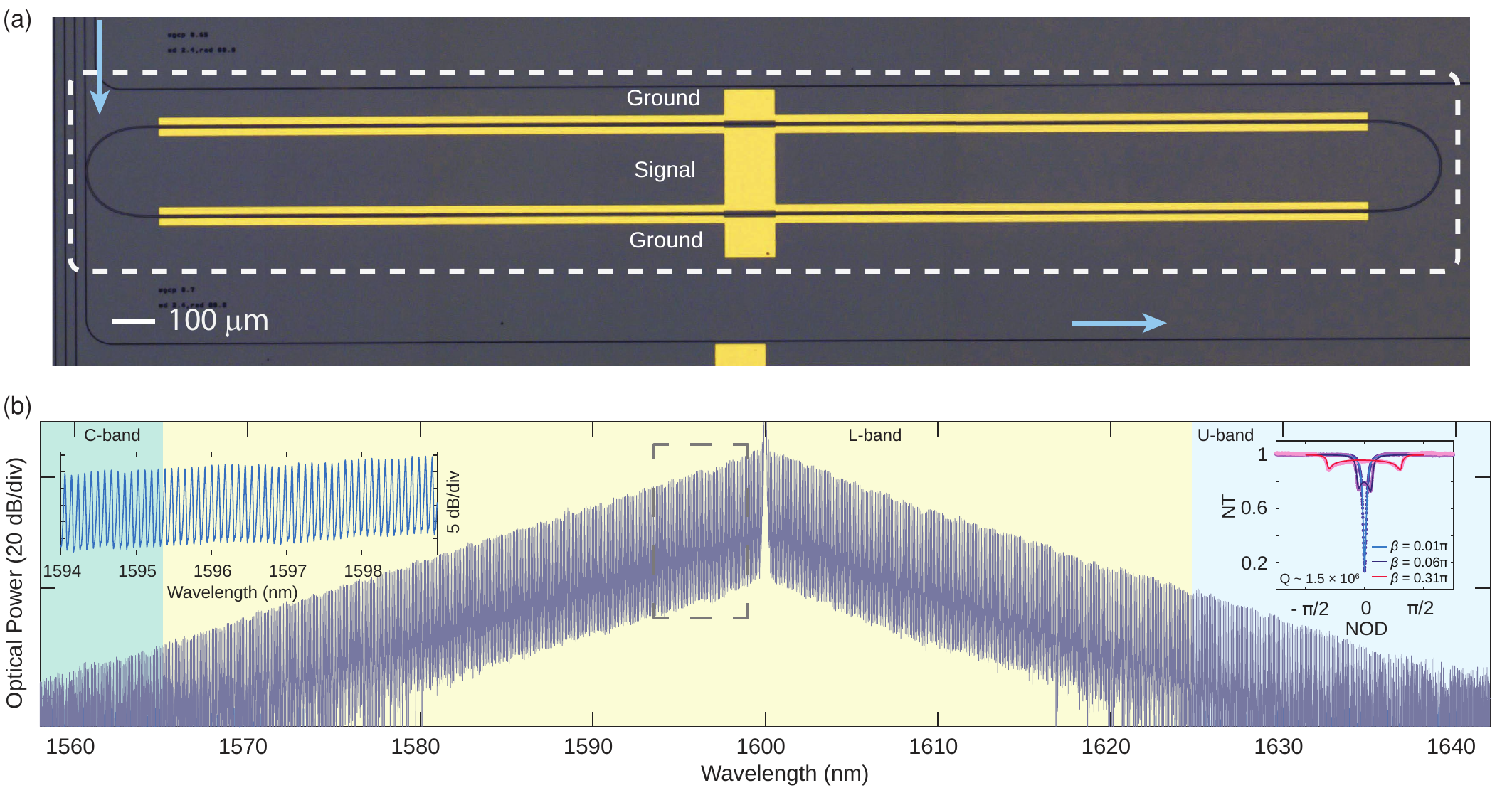}
	
	\caption{\label{fig2}\textbf{Integrated electro-optic comb generator.} \textbf{(a)} Micrograph of a fabricated lithium niobate microring resonator. The black lines are etched optical waveguides and the yellow regions are gold microelectrodes (see Supplementary). The gold electrodes are driven such that the phase shifts in the two sides of the microresonator are opposite, which is required to break the symmetry of different azimuthal order optical modes, enabling efficient frequency conversion. b, Measured output spectrum of the electro-optic comb generated from the microring resonator, demonstrating $>$ 80 nm bandwidth and more than 900 comb lines with a slope of 1 dB/nm. The input optical power is 2 mW and the microwave peak driving amplitude is $V_p$ = 10 V. Note that the signal-to-noise-ratio of the comb lines exceeds 40 dB but is limited by the noise floor and resolution of the optical spectrum analyzer. Insets: left, magnified view of several comb lines showing a line-to-line power variation of $\sim$ 0.1 dB. Right, measured transmission spectrum for several different modulation indices $\beta$. When the modulation is turned on, the optical resonance is broadened by twice the modulation index. This behaviour is predicted well by the round-trip phase model (see Supplementary). NT: normalized transmission. NOD: normalized optical detuning.}
	
\end{figure*}

Here we overcome these limitations of traditional discrete-component-based implementations by monolithically integrating an EO comb generator on a thin-film lithium niobate (LN) nanophotonic platform \cite{zhang_monolithic_2017-1,wang_nanophotonic_2017}. Leveraging the large $\chi^{(2)}$ nonlinearity, strong microwave and optical field overlap, and ultra-low loss optical waveguides enabled by this platform, we demonstrate integrated EO combs with performance superior to bulk EO comb generators. Our devices feature nearly two orders of magnitudes increase in comb width compared to previous integrated EO combs based on InP and Si platforms, which are limited by high optical losses \cite{dupuis_inp-based_2012,demirtzioglou_frequency_2018}. 

We demonstrate an EO frequency comb with over 900 unique frequencies spaced by 10.43 GHz, spanning 80 nm over part of the telecommunication C-band, the entire L-band and part of the U-band (Fig.\ref{fig2}). Our comb generator uses a low-loss LN microring resonator with loaded $Q\sim 1.5$ million, which is integrated with microwave electrodes for efficient phase modulation via the strong second-order nonlinearity of LN ($r_{33} = 30$ pm/V) (Fig. 2a) \cite{zhang_ultra-high_2018}. Importantly, the tight confinement of the light (waveguide width = 1.4 $\mu$m) allows for gold electrodes to be placed only 3.3 $\mu$m away from the edge of the resonator, resulting in efficient microwave delivery to achieve strong phase modulation while not affecting the resonator $Q$ factor. The two microwave electrodes are driven so the top and bottom sections of the resonator experience opposite phase shifts, phase matching the modulation to the circulating optical field. The microresonator is modulated by an external microwave synthesizer with peak voltage $V_p$ = 10 V ($\beta=1.2\pi$) at a frequency near the resonator FSR, and the generated comb spectrum (Fig. \ref{fig2}b) is well predicted by theory (see Supplementary).  The comb spectrum has a slope of $\sim$ 1 dB/nm (Fig. \ref{fig2}b left inset), corresponding to less than 0.1 dB power variation between adjacent comb lines. The comb lines have greater than 40 dB signal-to-noise ratio (SNR) near the pump frequency, where the measurement is limited by the noise floor and 20 pm resolution bandwidth of the optical spectrum analyser (OSA). 
\begin{figure*}[t!]
	\centering
	\includegraphics[angle=0,width=\textwidth]{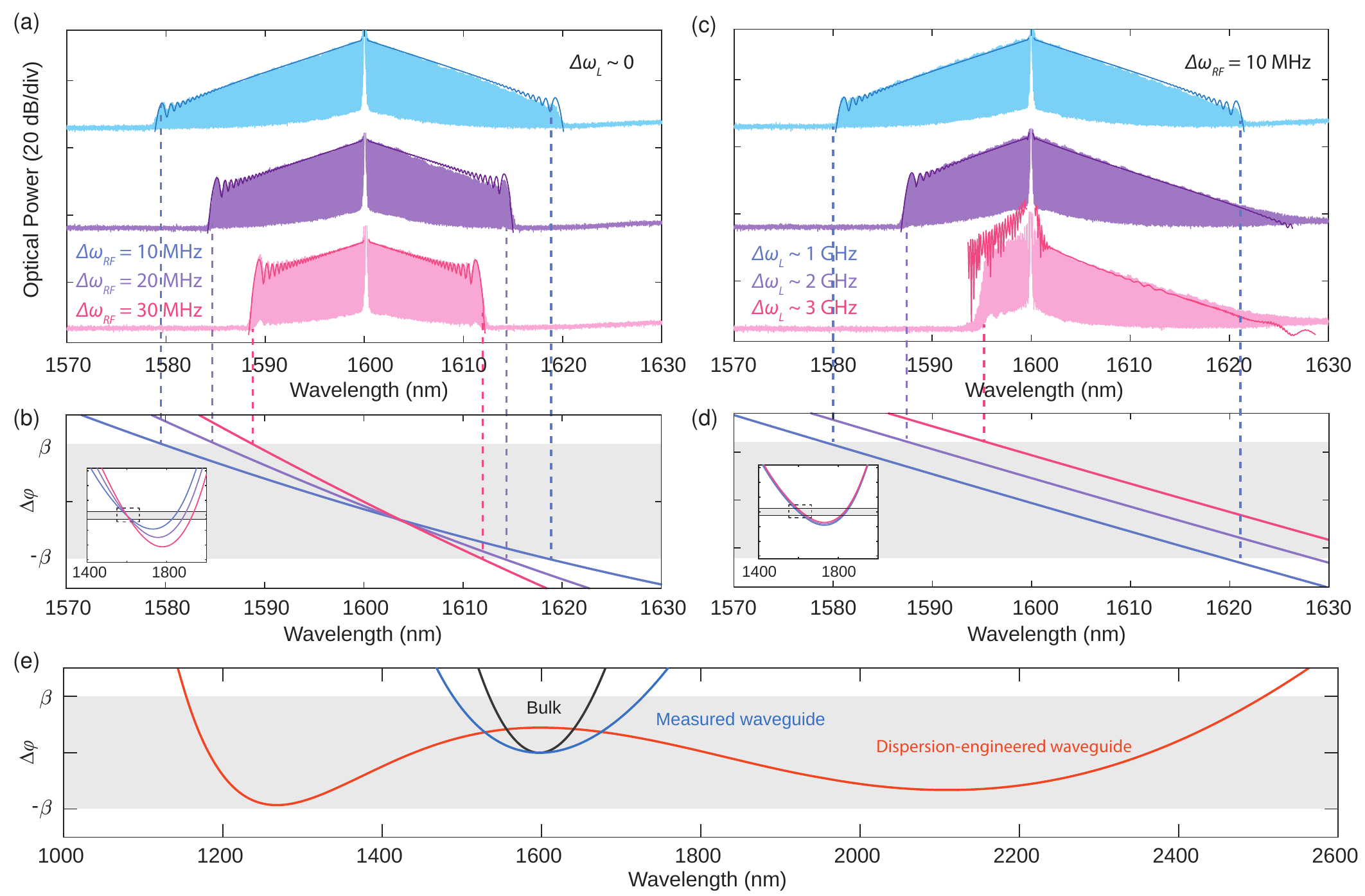}
	
	\caption{\label{fig3}\textbf{Controllability of the electro-optic comb spectrum.} \textbf{(a)} Measured electro-optic (EO) comb output spectrum for various values of modulation frequency detuning from the resonator free spectral range. Numerical simulation of the comb envelopes (dark lines, see Supplementary) match the measured spectra. \textbf{(b)} Calculated round-trip phase versus wavelength for the modulation frequency detuning values in (a). The light gray shaded region highlights the constructive interference condition region beyond which EO comb generation is suppressed. Insets show a zoomed-out view of the round-trip phase vs. wavelength. The calculated cut-off frequency matches well with experimental data, as shown by the dashed lines extending to (a). \textbf{(c,d)} Measured and simulated comb spectrum and round-trip phase versus wavelength in presence of both optical and microwave detuning. Different comb shapes, such as a single-sided EO comb can be generated. \textbf{(e)} Simulated round-trip phase versus wavelength for traditional bulk devices (black), the measured integrated device (blue), and dispersion-engineered integrated devices (orange). The simulations demonstrate that integrated EO combs can achieve larger dispersion-limited bandwidths than devices based on bulk crystals and dispersion engineering can enable octave-spanning EO combs. }
\end{figure*}

We develop a theoretical model to quantify the fundamental limits of the wide spanning EO combs generated on our integrated platform. Traditional EO comb span is limited to a narrow width by a combination of weak microwave modulation strength and native material dispersion, which hinders the constructive interference needed for cascaded frequency conversion to generate comb lines far from the pump frequency \cite{kourogi_limit_1995}. In contrast, the integrated EO comb generators feature large modulation strength and the ability to engineer dispersion, which enables broader EO comb generation. To understand such limitations, we look at the resonance condition for a comb line at optical frequency $\omega_q$. In a traditional static resonator, the round-trip constructive interference condition is $|\Delta\phi_q|<2l$, where $\Delta\phi_q=\omega_q T- 2\pi N$ is the accumulated round-trip phase, $T$ is the round-trip time and $N$ is the number of optical cycles per round-trip (chosen to minimize $|\Delta\phi_q|$). For optical frequencies that satisfy this condition, the optical field interferes constructively within the resonator. When the resonator length is modulated, as in an EO comb generator, the static resonance condition is modified into a dynamic one, where constructive interference occurs periodically at the microwave modulation frequency $\omega_m$ inside the resonator (i.e., $|\Delta\phi_q+\beta \textrm{sin}⁡\omega_mt|<2l$). Any frequency that does not satisfy this dynamic resonance condition will halt the frequency conversion process, thus limiting the comb width. This condition is reflected in the measured transmission spectrum of a microring resonator under microwave modulation (Fig. \ref{fig2}b right inset). With no microwave modulation ($\beta\sim 0$), the transmission spectrum exhibits a Lorentzian shape. By contrast, when the electrodes are strongly modulated (large $\beta$), the half-width at half-maximum of the transmission spectrum broadened by a factor of approximately $\beta$, confirming that the tolerable absolute accumulated phase $\Delta\phi$ is increased to $\beta$. It is therefore clear that it is the strong phase modulation achieved in integrated EO comb generator allowed for the continued cascade of phase modulation even in the presence of dispersion.

To verify the round-trip phase model experimentally, we detune the optical and microwave frequencies to generate different comb shapes and widths. By increasing the microwave detuning up to 30 MHz (Fig. \ref{fig3}a), we observe significant reduction in the comb frequency span, which is predicted well by the round-trip phase model (Fig. \ref{fig3}b). Any frequency components having total accumulated phases larger than $\beta$ cannot resonate, thus limiting the comb bandwidth. Taking advantage of this well understood dynamic resonance condition, we can generate asymmetric combs by appropriately choosing the optical and microwave detuning (Fig. \ref{fig3}c, d). EO combs driven off resonance, such as this one, could be used as low-noise sources for optical communications due to the noise-filtering properties of the optical resonator \cite{kim_cavity-induced_2017} (see Supplementary).

The ability to engineer the dispersion of integrated EO comb generators combined with the strong drive we can deliver can increase the achievable comb bandwidth up to a full octave. Traditionally, the span of EO comb generators is restricted by the dispersion of bulk materials, whereas EO comb generators that tightly confine light in optical waveguides enable fine tuning of  dispersion. For example, with a higher microwave modulation frequency of 50 GHz, high optical pump power (currently only 2 mW in our experiment), and a low-dispersion LN rib waveguide resonator, the round-trip phase model predicts the generation of an EO comb spanning over an octave (Fig. \ref{fig3}e and supplementary). 

\begin{figure*}
	\centering
	\includegraphics[angle=0,width=\textwidth]{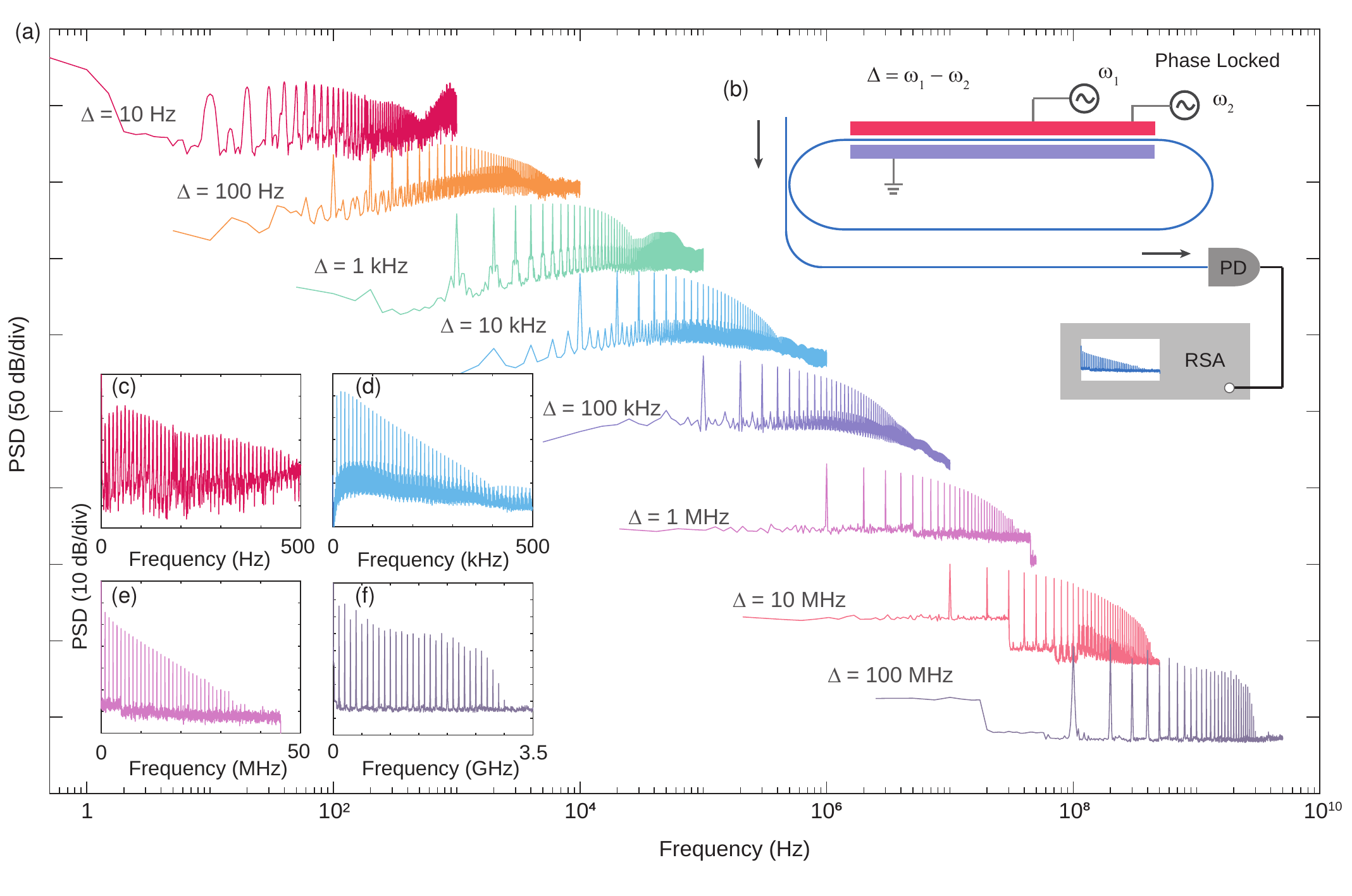}
	\caption{\label{fig4}\textbf{Dual-tone electro-optic comb generation} \textbf{(a)} Demonstration of coherent beating of the electro-optic (EO) comb. The measured beat-note power spectral density (PSD) is shown on a logarithmic scale to highlight the flexibility in control of the EO comb spacing over seven orders of magnitude, from 10 Hz to 100 MHz. \textbf{(b)} Experimental setup. The EO comb generator is driven by a superposition of two phase-locked microwave signals with various values of frequency offset $\Delta$. The optical output is detected by a fast photodiode (PD), and the beat notes are detected by a radio-frequency spectrum analyzer (RSA). \textbf{(c-f)} Magnification of the individual beat notes for various comb spacings on a linear frequency scale. This measurement, which demonstrates frequency components well beyond the static resonator bandwidth, confirms that phase modulation changes the resonance condition to tolerate large microwave detuning. Additionally, this measurement demonstrates the extreme flexibility in comb frequency spacing for practical applications such as dual-comb spectroscopy or comb-based ranging.}
\end{figure*}

Perhaps the most attractive properties of EO comb generators are their excellent configurability and stability. Leveraging the high tolerance to detuning the modulation frequency from the resonator FSR, we drive the microresonator electrodes with two phase-locked microwave sources at various frequency offsets, spanning over seven orders of magnitude, ranging from 10 Hz to over 100 MHz. The output of the comb generator is then connected to a high-speed photodetector, allowing observation of coherent beating between comb lines (Fig. \ref{fig4}). We note that this dual-driven EO comb contains frequency components far beyond the static resonance bandwidth (120 MHz) owing to the strong phase modulation. The ability to vary the frequency spacing of resonator-based EO combs over seven orders of magnitude, is in stark contrast with Kerr-based combs, whose frequency offset is predetermined by the fabricated resonator dimensions \cite{dutt_-chip_2018}. This flexibility in comb drive frequencies may enable applications requiring reconfigurable dynamic range, such as dual-comb-based optical ranging \cite{trocha_ultrafast_2018,suh_microresonator_2016} and spectroscopy \cite{coddington_dual-comb_2016,suh_microresonator_2016,dutt_-chip_2018,bernhardt_cavity-enhanced_2010}. Two independent microresonators can be integrated onto the same LN chip with high fabrication tolerance to avoid potential aliasing of the comb lines.

Our work using high-$Q$ microring resonators and highly confined optical waveguides for EO comb generation is the first step towards a new generation of integrated EO comb sources. Based on our demonstration of an EO comb that is almost two orders of magnitude larger than prior integrated EO combs, dispersion engineering and high frequency modulation can soon enable efficient octave-spanning EO comb generators.  Importantly, the approach demonstrated here can be used to realize EO combs all over the LN transparency window, including visible and near-IR, simultaneously. With the added ability to integrate filters and resonators adjacent or inside EO comb generators on the same chip, comb line power and hence SNR can be further increased by nearly 20 dB \cite{kourogi_coupled-cavity_1996}. These added components, including additional filters and modulators, enables application-specific EO comb photonic circuits. Our approach allows for complex EO circuits to be integrated on the same chip and thus has the potential to transform microresonator frequency comb applications. For example, high-performance EO combs featuring high power and flat combs could enable Tb/s optical communications links that rely on stable, low-noise combs as sources for high capacity wavelength-division multiplexed systems on a single chip \cite{marin-palomo_microresonator-based_2017}. Furthermore, the EO comb generator demonstrated in this work provides many stable coherent optical frequencies with electrically adjustable frequency spacing, paving the way for efficient dual-comb spectroscopy \cite{coddington_dual-comb_2016,bernhardt_cavity-enhanced_2010,dutt_-chip_2018,suh_microresonator_2016} on a chip or highly-reconfigurable comb-based ranging \cite{suh_soliton_2018,trocha_ultrafast_2018}.

This work is supported by: National Science Foundation award numbers ECCS-1609549, ECCS-1740291 E2CDA, ECCS-1740296 E2CDA, DMR-1231319; the Harvard University Office of Technology Development, Physical Sciences and Engineering Accelerator Award; and Facebook, Inc. Device fabrication is performed at the Harvard University Center for Nanoscale Systems, a member of the National Nanotechnology Coordinated Infrastructure Network, which is supported by the National Science Foundation under award number ECCS-1541959.

\bibliography{reference}

\pagebreak
\onecolumngrid

\begin{center}
	\textbf{\large Supplemental Information}
\end{center}

%Setup counters for supplementary
\setcounter{equation}{0}
\setcounter{figure}{0}
\setcounter{table}{0}
\setcounter{page}{1}
\makeatletter
\renewcommand{\theequation}{S\arabic{equation}}
\renewcommand{\thefigure}{S\arabic{figure}}
\renewcommand{\bibnumfmt}[1]{[S#1]}
\renewcommand{\citenumfont}[1]{S#1}
\section{Summary of Fabrication and Measurements Methods}
\subsection{Fabrication details} 
 All devices are fabricated on x-cut single crystalline thin-film lithium niobate (LN) wafers (NANOLN). The wafer stack consists of a 600 nm thin-film LN layer, a 2 $\mu$m thermally grown SiO$_2$ layer and a 500 $\mu$m silicon handle layer. Standard electron-beam (e-beam) lithography is used to pattern optical waveguide and micro-racetrack resonators. The patterns are then transferred into the LN layer using argon (Ar$^+$) plasma etching in an inductively coupled plasma reactive ion etching (ICP-RIE) tool13. The etch depth is 350 nm, leaving a 250 nm thick LN slab behind, which enables efficient electric field penetration into the waveguide core. Gold contact patterns are then created using aligned e-beam lithography, and the metal is transferred using e-beam evaporation methods and lift-off processes. The chip is then diced and the facet is polished for end-fire optical coupling. 

\subsection{Microwave driving circuitry }
The 10 GHz microwave drive signal is generated by a radio-frequency (RF) synthesizer and amplified by an electrical power amplifier. The amplified electrical signal is passed through a microwave circulator and delivered to the microelectrodes. As the microelectrodes represent a capacitive load, most of the electrical driving signal is reflected back to the circulator and terminated at the circulator output by a 50-$\Omega$ load. 

In the dual-drive EO comb generation experiment, two RF synthesizers are phase-locked via a common 10 MHz clock and are free to operate at different frequencies. The two sinusoidal microwave signals are power balanced and combined using an RF power splitter and passed through the amplifier-circulator circuitry described previously. 

\subsection{Optical characterization and detection}
Light from a tunable laser (SANTEC TS510) is launched into, and the comb output is collected from, the LN waveguides by a pair of lensed optical fibers. The output comb is passed to an optical spectrum analyser OSA having a minimum resolution of 20 pm. This finite resolution accounts for the limited signal-to-noise ratio observed in Fig 2b ($\sim$ 20 dB). The shot-noise-limited signal-to-noise ratio is much higher, as the comb shot noise lies below the OSA noise floor. Although the measurement in the paper is chosen to center at 1600 nm, the frequency comb center wavelength can be flexibly chosen between 1500 nm to 1680 nm of the tunable laser’s range without affecting much of the generated comb width.

In the dual-drive EO comb measurements, the modulated light is passed to a fast photodetector (New Focus 1544A) and the resulting electrical signal is sent to a RF spectrum analyzer to record the beating in the RF domain.  

\subsection{Measurement and calculation of resonator parameters }
As demonstrated by Equation (4) of the Supplementary Information, there are four resonator parameters that fully characterize the EO comb spectrum: the internal round-trip transmission coefficient α, the power coupling coefficient k, the coupler insertion loss of the coupler γ, and the phase modulation index β. Finding each of these four parameters by fitting to the comb spectrum of Equation (4) is difficult because the output comb can be fully determined by a subset of these independent parameters (e.g., increasing the modulation index has the same effect as decreasing the loss in the resonator). Instead, each of the parameters must be measured separately. 

We find $\alpha$ and $\kappa$ by measuring the total transmitted power without phase modulation (Figure 2b right inset). By fitting to the expected transmission of an all-pass ring resonator, we find $Q=1.5\times10^6, \alpha=0.95$ and $\kappa=0.027$. Then we perform a grid search optimization for $\gamma$ and $\beta$ comparing the measured output spectrum (Fig \ref{fig2}b) with the spectrum determined from the output time-domain electric field of Equation (3) of the Supplementary Information. We find a best fit for $\gamma=-0.004$ dB and $\beta=1.2 \pi$, where the average difference between experimental and theoretical comb line power is 0.6 dB.

The output power transmission for nonzero modulation indices (Fig. \ref{fig2}b right inset) is calculated by sampling the output electric field with Equation (3) of the Supplementary Information and averaging the power over more than 100 modulation periods. 

\subsection{Dispersion simulations in thin-film LN waveguides}
The dispersion of the waveguide is simulated using finite-domain-time-difference methods (LUMERICAL). The simulation accounts for the LN material anisotropy and the finite waveguide etching angle (around $70^\circ$ from horizontal). The round-trip phase of the light inside the resonator is calculated by integrating the simulated group velocity dispersion twice to determine the total frequency-dependent phase-shift. We find that for an air-cladded waveguide with a 600 nm thin-film LN layer, 350 nm etch depth and 1.5 $\mu$m waveguide width, a comb spanning $\sim$ 1.2 octave can be generated, as shown in Fig \ref{fig3}e.

\section{Canonical EO Comb Generator Design}

\subsection{Resonant Operation}

A canonical waveguide-based comb generator is shown in Figure 1c of the main text. A single-frequency input with electric field $E_{in}(t)=\hat{E}_{in} e^{i \omega_0 t}$  is coupled, with power coupling coefficient $k$ and insertion loss $\gamma$, to a resonator having round trip time $T$ at center frequency $\omega_0$ and round trip power loss $\alpha$. The resonator contains a phase modulator driven with modulation index $\beta$ and frequency $\omega_m$. The output electric field is ~\cite{ho_optical_1993}:

\begin{equation}
\begin{aligned}
\label{eq:E_out1}
E_{out} = \sqrt{(1-\gamma)(1-k)} E_{in}(t) - k \sqrt{\frac{1-\gamma}{1-k}} \sum_{n=1}^{\infty} r^n e^{-i \beta F_n(\omega_m t)} E_{in} (t -n T),
\end{aligned}
\end{equation}

where $r = \sqrt{(1-\gamma)(1-k) \alpha}$ is the round trip electric field transmission and $ F_n(\omega_m t)= \sum_{i=1}^n \sin{\omega_m(t-iT)}$ is the modulator coherence function. The parameter $l = 1-r$, corresponding to the round-trip electric field loss, is used in the main text for simplicity. When the optical carrier is resonant ($\omega_0 T = 2 \pi m_1$) and the microwave drive signal is resonant ($\omega_m T = 2 \pi m_2$), the modulator coherence function becomes $F_n (\omega_m t) = n \sin{\omega_m(t-iT)}$ and the output electric field can be simplified to

\begin{equation}
\begin{aligned}
\label{eq:E_out2}
E_{out}(t) = \bigg[ \sqrt{(1-\gamma)(1-k)} - k \sqrt{\frac{1-\gamma}{1-k}} \frac{r e^{-i \beta \sin{\omega_m t}}}{1 - r e^{-i \beta \sin{\omega_m t}}} \bigg] E_{in}(t).
\end{aligned}
\end{equation}

This output electric field corresponds to an optical frequency comb spaced at the modulation frequency. The power in the $q$th comb line away from the center frequency can be found by rewriting Equation \ref{eq:E_out1} as

\begin{equation}
\begin{aligned}
\label{eq:E_out3}
E_{out}(t) = \sqrt{(1-\gamma)(1-k)} \hat{E}_{in} e^{i \omega_0 t} - k \sqrt{ \frac{1-\gamma}{1-k} } \hat{E}_{in} \sum_{q = -\infty}^{\infty} e^{i(\omega_0+q \omega_m)t} \sum_{n=1}^{\infty} r^n J_q(\beta n),
\end{aligned}
\end{equation}

where $J_q$ is the $q$th order Bessel function of the first kind. The power of the $q$th (nonzero) comb line is then

\begin{equation}
\begin{aligned}
\label{eq:Pq_r}
P_q = k^2 \frac{1-\gamma}{1-k} P_{in} \bigg| \sum_{n = 1}^{\infty} r^n J_q(\beta n) \bigg|^2 .
\end{aligned}
\end{equation}

~\cite{kourogi_wide-span_1993} found an approximation for the power of the $q$th comb as $P_q \propto e^{-\frac{|q|(1-r)}{\beta}}$.

\subsection{Nonresonant Operation}

In the presence of optical and microwave detuning from resonance, the comb spectrum can still be calculated. When the optical carrier is off resonance, the total round-trip phase is $\omega_0 T = 2 \pi m_1+ \phi_{opt}$. Similarly, when the microwave carrier is off resonance the total round-trip phase is $\omega_m T = 2 \pi m_2+ \phi_{micro}$. Using these expressions in Equation~\ref{eq:E_out1}, we can find the following expression for the power in the $q$th comb line:

\begin{equation}
\begin{aligned}
\label{eq:Pq_nr}
P_q = k^2 \frac{1-\gamma}{1-k} P_{in} \bigg| \sum_{p = -\infty}^{\infty} \sum_{n = 1}^{\infty} (r e^{i \phi_{opt}})^n e^{i p \frac{\pi}{2}} J_{q-p}(\beta_o(\phi_{micro},n)) J_p(\beta_e(\phi_{micro},n)) \bigg|^2 .
\end{aligned}
\end{equation}

The modified even and odd modulation indices ($\beta_e$ and $\beta_o$, respectively) are

\begin{equation}
\begin{aligned}
\label{eq:beta_e}
\beta_e(\phi_{micro},n) = \beta \bigg[ \frac{1}{2} \cot{\phi_{micro}/2} - \frac{\cos{(n + \frac{1}{2}) \phi_{micro}}}{2 \sin{\phi_{micro}/2}} \bigg]
\end{aligned}
\end{equation}

\begin{equation}
\begin{aligned}
\label{eq:beta_o}
\beta_o(\phi_{micro},n) = \beta \bigg[ -\frac{1}{2} + \frac{ \sin{(n + \frac{1}{2}) \phi_{micro}} }{2 \sin{\phi_{micro}/2}} \bigg].
\end{aligned}
\end{equation}

It is clear here that in the regime of low optical detuning, the slope of the comb decreases by a factor of $\cos{(\phi_{opt})}$. This effect has been studied and reported in~\cite{saitoh_modulation_1998}. The effect of microwave detuning is harder to visualize, but results in a destructive interference condition for large values of $q$ in Equation~\ref{eq:Pq_nr}. This effect is demonstrated experimentally and theoretically in Figure 3a and 3b of the main text.

\subsection{Noise Properties}

The optical phase noise of the comb lines is important in applications that require high optical signal-to-noise ratios, such as high-capacity optical communications. It is well known that the optical phase noise contribution from the pump laser does not increase with increasing comb line index $q$ \cite{ho_optical_1993}. By contrast, the phase noise contribution from the microwave modulation signal increases in power with comb line quadratically with $q$. This can be shown by modifying the modulator coherence function to include the effects of microwave modulation phase noise $\theta(t)$:

\begin{equation}
\begin{aligned}
\label{eq:F_n}
F_n(\omega_m t) = \sum_{i=1}^n \sin{\omega_m (t - i T + \theta(t - i T) )}.
\end{aligned}
\end{equation}

The output optical field can then be written as:

\begin{equation}
\begin{aligned}
\label{eq:E_out4}
E_{out}(t) = \sqrt{(1-\gamma)(1-k)} \hat{E}_{in} e^{i \omega_0 t} - k \sqrt{\frac{1-\gamma}{1-k}} \hat{E}_{in} \sum_{p = -\infty}^{\infty} \sum_{n = 1}^{\infty} r^n J_q(\beta n) e^{i(\omega_0 + q \omega_m)t + i q \theta(t)}.
\end{aligned}
\end{equation}

The phase noise amplitude increases linearly with increasing comb line index $q$, corresponding to a quadratic increase in phase noise power.

For applications that require few comb lines, this increase in microwave phase noise is often negligible because quartz crystal oscillators have very low phase noise. For applications requiring many comb lines, however, the effect of microwave phase noise may be noticeable. Recently, there has been experimental evidence of microwave phase noise suppression in EO comb generators \cite{kim_cavity_2016}, \cite{kim_cavity-induced_2017}. In these studies, the phase noise increase can be dramatically suppressed when the EO comb generator is driven off resonance, both optically and electrically. These experiments suggest that EO comb generators can generate low-noise comb lines over their entire dispersion-limited bandwidth. Integrated platforms, such as the one presented in the main text, enable additional filtering cavities and structures to be readily included in the resonator structure.

\section{Round-Trip Phase Model}

To include the effect of dispersion, we introduce a round-trip phase model. In particular, we consider the destructive interference that occurs due to the microwave detuning motivates a phase-based resonance approximation for the viable comb bandwidth. Previous analytical work~\cite{kourogi_limit_1995} provided a mathematical treatment of the dispersion limits of resonator-based EO comb generators. Here, as a complement to that work, we clarify the physical interpretation of the round-trip phase model and demonstrate its application to combs of arbitrary bandwidth within a given dispersion-limited window.

\subsection{Resonance Conditions}

The resonance condition of an optical frequency $\omega_q$ in a microresonator without EO modulation is $|\omega_q T - 2 \pi N | < 2 l$, where the total round-trip phase offset $\Delta \phi_q = \omega_q T - 2 \pi N$, $T=1/$FSR is the round-trip time and $N$ is the number of optical cycles per round-trip that ensures that $|\Delta \phi_q | < 2 \pi$. Frequency components outside of the resonance are attenuated by destructive interference, and thus do not resonate. When the resonance condition is satisfied, the optical fields constructively interfere inside the resonator at every time and spatial location.

In a resonator containing an EO phase modulator, the (now time-dependent) resonance condition becomes $|\Delta \phi_q + \beta \sin{2 \pi f_m t}| < 2l$, where $\beta$ is the modulation index and $f_m$ is the modulation frequency. Here, it is clear that the resonance condition can be satisfied for much larger round-trip phase offsets $\Delta \phi_q$ because within the round-trip resonator propagation time, the modulation term oscillates between negative and positive $\beta$ (i.e. $-\beta < \beta \sin{2 \pi f_m t} < \beta$). 

This effect may be understood by plotting the total transmission of the EO comb generator for various $\beta$, as shown in Figure 2b, right inset, of the main text. The transmission is calculated by averaging the output power of a time-domain representation of the electric field given in Equation~\ref{eq:E_out1}. The optical power output depends primarily on the interference between the input optical field and the optical field inside the resonator. As in a microresonator without EO modulation, the dips in the transmission spectrum correspond to a large built-up field inside the resonator. For various values of $\beta$, the width of the resonance increases, indicating that for large modulation indices, the resonance condition can be satisfied for various detuning values. As shown in Figure 2b, the amount of detuning is approximately equal to the modulation index $\beta$, as is predicted by the phase model when $\Delta \phi_q= \phi_{opt}$.

\subsection{Frequency-Dependent Round-Trip Phase}

We can now determine the contributions to the optical phase offset $\Delta \phi_q$ as a function of frequency. The optical phase offset, as discussed in the previous section, does not induce frequency-dependent phase shifts. However, microwave signal detuning and dispersion effects are frequency dependent.

Once the resonator has reached steady state, the output field is an EO comb spaced at the modulation frequency $f_m$, such that the $q$th comb line frequency is $f_q = f_0+ q f_m$. A mismatch between the microwave frequency and the resonator free spectral range results in a frequency-dependent phase offset $\phi_{micro}(q) = 2 \pi q $(FSR)$ T$. 

For an arbitrary dispersion profile, it is possible to find the frequency-dependent phase offset by integrating the group velocity dispersion profile of the waveguide. However, if the dispersion is approximately linear with frequency, the dispersion-related phase offset is $\Delta \phi_{disp}(q) = 2 \pi (q f_m)^2 \beta_2 L $ where $\beta_2 L$ is the round-trip group velocity dispersion in ps/nm.

To first order we then have a model for the total phase offset as a function of frequency, $\Delta \phi_q = \Delta \phi_{opt}+ \Delta \phi_{micro}(q) + \Delta \phi_{disp}(q)$. In fact, this model agrees exactly with the analytical model for the output comb shape developed in~\cite{kourogi_wide-span_1993}. In the case of maximum comb bandwidth, corresponding to zero microwave detuning and optical detuning satisfying $\phi_{opt} + \beta = 0$, the maximum dispersion-limited bandwidth is $ \Delta f_{comb} = \frac{1}{\pi} \sqrt{\frac{2 \beta}{\beta_2 L}} $, agreeing with~\cite{kourogi_limit_1995} up to a factor of $\sqrt{2}$ due to the difference in FSR of a Fabry-Pérot resonator and ring resonator of identical length.

Using this model, it is a straightforward optimization problem to start with the frequency-dependent round-trip resonance condition and alter the optical and microwave detuning so that the resonance condition is satisfied only for a desired frequency region, as is done to demonstrate the one-sided comb in Figure 3c of the main text.

\end{document}